\documentclass[conference]{IEEEtran}

\IEEEoverridecommandlockouts
\usepackage{amsmath}
\usepackage{amsfonts}
\usepackage{amssymb}

\usepackage{graphicx}
\usepackage{psfrag}
\usepackage{epsfig}
\usepackage{tikz}
\usetikzlibrary{%
  arrows,%
  shapes.misc,
  shapes.arrows,%
  chains,%
  matrix,%
  positioning,
  scopes,%
  decorations.pathmorphing,
  shadows%
}
\usepackage{pgfplots}

\ifCLASSINFOpdf
\else
\fi
\begin{document}
%
\title{Chromatic Dispersion Compensation Using Filter Bank Based Complex-Valued All-Pass Filter}

\author{\IEEEauthorblockN{Jawad Munir, Amine Mezghani, Haris Khawar, Israa Slim and Josef A. Nossek}

\IEEEauthorblockA{Institute for Circuit Theory and Signal Processing, 
Technical University Munich, Germany\\
Email: $\{$jawad.munir, amine.mezghani, haris.khawar, israa.slim, josef.a.nossek$\}$@tum.de}
}


%


\maketitle

\begin{abstract}
A long-haul transmission of $100$ Gb/s without optical chromatic-dispersion (CD) compensation provides a range of benefits regarding cost effectiveness, 
power budget, and nonlinearity tolerance. The channel memory is largely dominated by CD in this case with an intersymbol-interference spread of more
than $100$ symbol durations. In this paper, we propose CD equalization technique based on nonmaximally decimated discrete Fourier transform (NMDFT) filter bank (FB) with 
non-trivial prototype filter and complex-valued infinite impulse response (IIR) all-pass filter per sub-band. The design of the sub-band IIR all-pass filter 
is based on minimizing the mean square error (MSE) in group delay and phase cost functions in an optimization framework. Necessary conditions are derived
and incorporated in a multi-step and multi-band optimization framework to ensure the stability of the resulting IIR filter. It is shown that the complexity 
of the proposed method grows logarithmically with the channel memory, therefore, larger CD values can be tolerated with our approach.
\end{abstract}


%
\IEEEpeerreviewmaketitle


\section{Introduction}

The performance of fiber optic links in long haul, metro and enterprise networks became limited by chromatic dispersion causing a short optical pulse to broaden as it
travels along the fiber leading to intersymbol interference (ISI). To mitigate the effects of dispersion, optical systems include some form of CD compensation either 
electronically, optically or digitally. Digital equalization is attractive because it is less expensive, flexible and robust to varying channel conditions compared 
to electronic and optical. 

CD has a time-invariant transfer function and only effects the phase of the input signal, i.e., it exhibits an all-pass behavior. In \cite{Goldfarb2007}, G. Goldfarb and G. Li suggested dispersion compensation (DC) using time domain equalizer (TDE) technique based 
on infinite impulse response (IIR) all-pass filter. Their methodology requires Hilbert transformer and time reversal operation to design real-coefficients 
IIR filter separately for real and imaginary part of the transmitted signals. This technique was improved by J. Munir et al. \cite{Munir2014} in which an 
optimization framework was presented to design stable complex-valued IIR all-pass filter. Computational complexity of this scheme is less compared to \cite{Goldfarb2007} 
as it doesn't require the use of Hilbert transformer. However TDE schemes are not feasible for very long optical link as their complexity increases linearly with 
the channel memory. 

Frequency domain equalizers (FDEs) based on fast Fourier transform (FFT) have become the most appealing scheme
for CD compensation due to the low computational complexity for large dispersion and the wide applicability for different fiber distances \cite{Ishihara2011}. In state-of-the-art FDE CD compensation design, the size of the FFT to realize fast linear convolution is governed by the specification of the maximum channel memory
length with two-fold oversampling and $50\%$ block overlap \cite{Ishihara2008}. Nevertheless, the overlap-save method can be regarded not just as fast convolution method but also 
as a FB structure with trivial prototype filters and the equalization is done per sub-band, opening the way for more sophisticated sub-band processing \cite{Slim2012}. Based on this observation, we propose CD equalization technique based on NMDFT FB with non-trivial prototype filter and complex-valued IIR all-pass filter per sub-band.

The main contribution of the paper is the presentation of an optimization framework to design stable sub-band complex-valued IIR all-pass filter for CD equalization. 
Root-raised-cosine (RRC) filter is chosen as a non-trivial prototype filter and optimal filter length and roll-off to achieve certain bit-error-rate (BER) 
are also investigated. It is shown that the complexity of our approach increases logarithmically with the channel memory. 
The paper is organized as follows. Section \ref{Channel Model} introduces the channel transfer function for CD. Section \ref{Equalizer Design} describes the equalizer 
description based on IIR filtering and its complexity analysis. FB based sub-band equalizer design is presented in Section \ref{Filter Bank based Complex Valued Sub-band All-Pass Equalizer} and 
necessary condition for stability and filter order of individual sub-bands are derived in Section \ref{sec:Complex Value Sub-band All-Pass Equalizer Design}. 
Two design criteria presented in an optimization framework \cite{Munir2014} are modified in Section \ref{sec:Design Criteria For the Sub-Band All-Pass Filters} for sub-band equalizer design. 
Comlexity analysis of the proposed scheme is derived in Section \ref{sec:Complexity Analysis}. 
Simulation results comparing IIR and FB based IIR filtering are presented in Section \ref{sec:Simulations}. Conclusions are given in Section \ref{sec:Conclusions}.
\section{Channel Model}
\label{Channel Model}
The low-pass equivalent model of CD channel of a single mode fiber of length $L$ can be written as

\begin{equation}
H_{\textrm{CD}}(\Omega)=\exp\left(-j\cdot\frac{\lambda_{0}^{2}}{4 \pi c}\cdot D\cdot L\cdot\Omega^{2}\right)\textrm{ ,}
\end{equation}
where $\Omega$, $\lambda_0$, $D$ and $c$ are baseband radial frequency, operating wavelength, fiber dispersion parameter and speed of light, respectively. 
If we sample the signal by sampling frequency $B$ Hz, then the equivalent model in discrete domain can be represented as
\begin{equation}
H_{\textrm{CD}}(\omega)=\exp(-j\cdot\alpha\cdot\omega^{2})\textrm{ ,}\label{eq: CD Transfer Function}
\end{equation}
where $\alpha=\lambda_{0}^{2}\cdot B^{2}\cdot D\cdot L /\left(4\pi c\right)$ and $\omega\in\left[\textrm{-}\pi,\pi\right)$. It is clear from 
(\ref{eq: CD Transfer Function}) that the CD channel has an all-pass characteristic, i.e., it only changes the phase of the input signal. 

\section{Equalizer Design}
\label{Equalizer Design}
\begin{figure}
\centering
\psfrag{M}[][][1.3]{$\frac{M}{2}$}
\psfrag{ra}{\tiny $  {\rm Re} \{\! a_i \!  \}$}
\psfrag{ia}{\tiny  ${\rm Im} \{ \! a_i \!  \}$}
\psfrag{Rev}{\tiny $  {\rm Re} \{\!v [n]\!  \}$}
\psfrag{Imv}{\tiny  ${\rm Im} \{ \! v[n] \!  \}$}
\psfrag{Rey}{\tiny $  {\rm Re} \{\! y [n] \!  \}$}
\psfrag{Imy}{\tiny  ${\rm Im} \{ \! y [n] \!  \}$}
\includegraphics[scale=0.8]{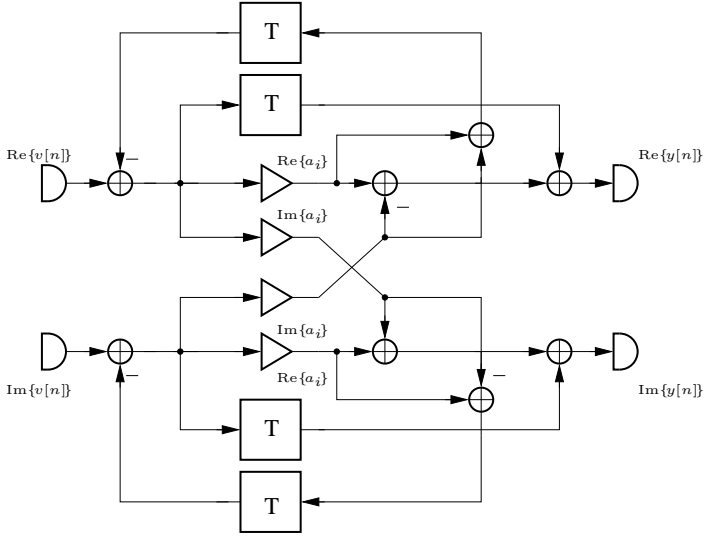}
\caption{\label{fig:All-pass efficient implementation} Efficient implementation of 1st order IIR all-pass section.}
\end{figure}

The transfer function of an ideal equalizer to compensate CD channel is obtained by taking the inverse of (\ref{eq: CD Transfer Function}) 
\begin{equation}
G_{\textrm{Ideal}}(\omega)=\exp(+j\cdot\alpha\cdot\omega^{2})\textrm{ .}\label{eq: Ideal Eq. Transfer Function}
\end{equation}
Since the CD channel exhibits an all-pass behavior, therefore, the natural choice to equalize it by the cascade of $N_{\textrm{IIR}}$ first order IIR all-pass sections of the form
\begin{equation}
G_{\textrm{IIR}}\left(z\right)=\prod_{i=1}^{N_{\textrm{IIR}}}\frac{-\rho_{i}e^{-j\theta_{i}}+z^{-1}}{1-\rho_{i}e^{j\theta_{i}}\cdot z^{-1}}\textrm{ ,}\label{eq:Cascade All-Pass Polar Form}
\end{equation}
where $\rho_i$ and $\theta_i$ are the radius and angle of the $i^{th}$ pole location in the complex $z$-plane. Each first order section of complex valued IIR all-pass filter can be efficiently realized by four real multiplications as shown in Fig. \ref{fig:All-pass efficient implementation}. It was shown in \cite{Munir2014} that the required filter order is given as  
\begin{equation}
N_{\textrm{IIR}}=\left\lceil\left(\frac{\lambda_{0}^{2}}{2c}\right)\cdot D\cdot B^{2}\cdot L\right\rceil\textrm{.}\label{eq: Filter Order All-Pass}
\end{equation}
It can be seen from above relation that the $N_{\textrm{IIR}}$ increases linearly with the fiber length which is not feasible to equalize very long optical links. In order 
to extend this framework to equalize long optical channels, we present FB based complex-valued all-pass equalizer design.

\section{Filter Bank based Complex Valued Sub-band All-Pass Equalizer}
\label{Filter Bank based Complex Valued Sub-band All-Pass Equalizer}
\begin{figure}
\begin{center}

\psfrag{T}{$T$}
\psfrag{M2}{$\frac{M}{2}$}

\footnotesize \psfrag{G_0(z_1)}{ $G_{0}(z^{M/2})$}
\footnotesize \psfrag{G_1(z_1)}{ $G_{1}(z^{M/2})$}
\footnotesize \psfrag{G_M1(z_1)}{ $G_{M\!-\!1}(z^{M\!/\!2})$}

\psfrag{y_0[m]}{$y_{0}[m]$}
\psfrag{y_1[m]}{$y_{1}[m]$}
\footnotesize     \psfrag{y_M1[m]}{$y_{M\!-\!1}[m]$}

\psfrag{yt0}{$\tilde{y}_{0}[\ell]$}
\psfrag{yt1}{$\tilde{y}_{1}[\ell]$}
\footnotesize    \psfrag{ytM}{$\tilde{y}_{M\!-\!1}[\ell]$}

\psfrag{x[n]}{$x[n]$}

\includegraphics[width=8.5cm]{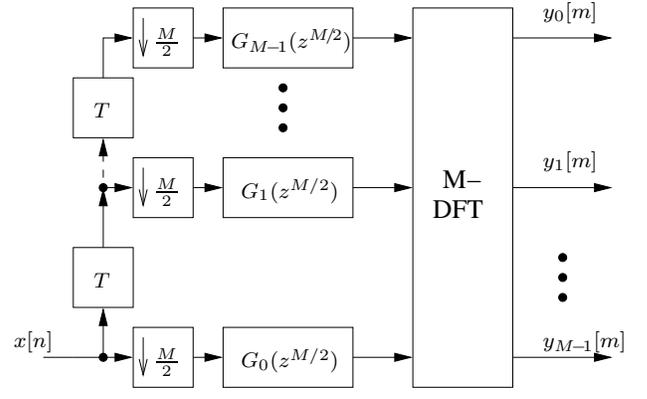}
\caption{Efficient structure of analysis FB of non-maximally decimated DFT FB: the serial discrete-time input signal $x[n]$ is first parallelized (with 50\% overlap) to get overlapping blocks that are filtered by the polyphase network and finally DFT transformed.}
\label{fig:afb_polyphase}
\end{center}
\end{figure}

\begin{figure}
\begin{center}

\psfrag{T}{$T$}
\psfrag{M2}{$\frac{M}{2}$}

\footnotesize \psfrag{G_0(z_1)}{ $\tilde{G}_{0}(z^{M/2})$}
\footnotesize \psfrag{G_1(z_1)}{\;\;  $\tilde{G}_{1}(z^{M/2})$}
\footnotesize \psfrag{G_M1(z_1)}{ $\tilde{G}_{M\!-\!1}\!(z^{M\!/\!2})$}

\psfrag{yhat0}{\;\;\;\;\;$\hat{y}_{0}[m]$}
\psfrag{yhat1}{\;\;\;\;\;$\hat{y}_{1}[m]$}
\footnotesize \psfrag{yhatM1}{\;\;$\hat{y}_{M\!-\!1}[m]$}

\psfrag{yt1}{$\tilde{y}_{0}[\ell]$}
\psfrag{yt2}{$\tilde{y}_{1}[\ell]$}
\footnotesize    \psfrag{ytM}{$\tilde{y}_{M\!-\!1}[\ell]$}

\psfrag{xhat}{$\hat{x}[n]$}

\includegraphics[width=8cm]{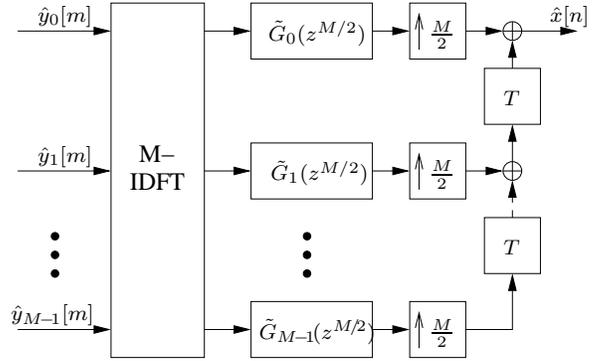}
\caption{Efficient structure of synthesis FB of non-maximally decimated DFT FB: the input block to the synthesis FB is IDFT transformed, then filtered by the polyphase network and finally serialized to give the output signal $\hat{x}[n]$.}
\vspace{-0.2cm}
\label{fig:sfb_polyphase}
\end{center}
\end{figure}

Fig. \ref{fig:afb_polyphase} and Fig. \ref{fig:sfb_polyphase} illustrates the efficient implementation of NMDFT filter bank structure \cite{Slim2013} with RRC filter as a non-trivial prototype filter of length $L_{\textrm{RRC}}=KM$ where $K$ is a positive integer number. The polyphase components of the RRC filter in the analysis filter bank (AFB) and synthesis filter bank (SFB) are 
$G_k(z)$ and $\tilde{G}_k(z),\textrm{ }k = 0,\cdots,M-1$ respectively. The basic idea is to divide the incoming signal into $M$ frequency bands after filtering and downsampling operation in AFB and design IIR all-pass CD equalizer for every sub-band. The complexity of the overall structure is reduced since each frequency band is operating at a lower rate $M/2$. The choice of $M$ is independent of channel memory which can be advantageous from hardware implementation perspective. 




\section{\label{sec:Complex Value Sub-band All-Pass Equalizer Design}Complex Valued Sub-band All-Pass Equalizer Design}

Transfer function of the sub-band IIR all-pass filter has the form 
\begin{equation}
G^{(k)}_{\textrm{IIR}}\left(z\right)=\prod_{i=1}^{N^{(k)}_{\textrm{IIR}}}\frac{-\rho^{(k)}_{i}e^{-j\theta^{(k)}_{i}}+z^{-1}}{1-\rho^{(k)}_{i}e^{j\theta^{(k)}_{i}}\cdot z^{-1}}\textrm{ , }k = 0,\cdots,M-1\textrm{.}\label{eq:Sub-band Cascade All-Pass Polar Form}
\end{equation}
The design strategy for each sub-band all-pass equalizer will be same as described in \cite{Munir2014}. But we consider it important to derive the
phase response and group delay as seen by every sub-band. Based on the group delay behavior, we will describe the stability criteria and the optimum number of stages for every sub-band.

\subsection{Phase Response And Group Delay }

Transfer function of the CD channel in the $k^{th}$ sub-band after filtering and downsampling operation is given by 
\begin{equation}
H_{\textrm{CD}}^{\left(k\right)}\left(\omega^{'}\right)=\exp(-j\cdot\alpha^{'}\cdot\left(\omega^{'}+k^{'}\cdot\pi\right)^{2})\textrm{,}\label{eq:Sub-Band CD Transfer Function}\end{equation}
where $\alpha^{'}=\alpha\cdot\left(\frac{2}{M}\right)^{2}$, $\omega^{'}\in\left[\textrm{-}\pi,\pi\right)$, $k^{'}=k\textrm{ for } k\leq\frac{M}{2}$ and $k^{'}=k-M\textrm{ for } k>\frac{M}{2}$. Transfer function of an ideal equalizer to compensate 
CD channel is obtained by taking the inverse of (\ref{eq:Sub-Band CD Transfer Function})
\begin{equation}
G_{\textrm{Ideal}}^{\left(k\right)}\left(\omega^{'}\right)=\exp(j\cdot\alpha^{'}\cdot\left(\omega^{'}+k^{'}\cdot\pi\right)^{2})\textrm{.}\end{equation}
Phase response and group delay for the $k^{th}$ sub-band of an ideal equalizer takes the following form
\begin{equation}
\phi_{\textrm{Ideal}}^{\left(k\right)}\left(\omega^{'}\right)=\alpha^{'}\cdot\left(\omega^{'}+k^{'}\cdot\pi\right)^{2}\textrm{.}
\label{eq: Sub-Band Phase Response Ideal Equalizer}\end{equation}
\begin{equation}
\tau_{\textrm{Ideal}}^{\left(k\right)}\left(\omega^{'}\right)=-2\cdot\alpha^{'}\cdot\left(\omega^{'}+k^{'}\cdot\pi\right)\textrm{.}\label{eq: Sub-Band Group Delay Ideal Equalizer}\end{equation}
Negative group delay implies an unstable all-pass filter \cite{Abel2006}, therefore, constant $\beta^{'}$ is added to the ideal group $\tau_{\textrm{Ideal}}^{\left(k\right)}\left(\omega^{'}\right)$ for all the sub-bands to make 
non-negative desired group delay function, i.e.,
\begin{equation}
\tau_{\textrm{Desired}}^{\left(k\right)}\left(\omega^{'}\right)=\tau_{\textrm{Ideal}}^{\left(k\right)}\left(\omega^{'}\right)+\beta^{'}\textrm{,}\label{eq: Sub-Band Desired Group Delay}
\end{equation}
where
\begin{equation}
\beta^{'}=\left\lceil -\tau_{\textrm{Ideal}}^{\left(\frac{M}{2}\right)}\left(\omega^{'}=0\right)\right\rceil=\left\lceil 2\cdot\alpha^{'}\cdot\frac{M}{2}\cdot\pi\right\rceil\textrm{.} 
\end{equation}
Desired phase takes the form
\begin{equation}
\phi_{\textrm{Desired}}^{\left(k\right)}\left(\omega^{'}\right)=\alpha^{'}\cdot\left(\omega^{'}+k^{'}\cdot\pi\right)^{2}-\beta^{'}\cdot\left(\omega^{'}+k^{'}\cdot\pi\right)+\phi_{0}^{\left(k\right)}\textrm{.}\label{eq:Sub-Band Desired Phase Response}
\end{equation}
Here $\phi_{0}^{\left(k\right)}$ is a constant of integration and it will be calculated in the optimization framework.
\subsection{Filter Order}
Filter order for the $k^{th}$ sub-band is obtained by integrating its desired group delay function (\ref{eq: Sub-Band Desired Group Delay})
\begin{equation}
\sum_{i=1}^{N^{(k)}_{\textrm{IIR}}}\int_{-\pi}^{\pi}\tau_{\textrm{IIR}_{i}}^{\left(k\right)}\left(\omega^{'}\right)d\omega^{'}=2\pi N^{(k)}_{\textrm{IIR}}\textrm{ .}
\end{equation}
But the integral of the desired group delay is given by
\begin{equation}
\int_{-\pi}^{\pi}\tau_{\textrm{Desired}}^{\left(k\right)}\left(\omega^{'}\right)\cdot d\omega^{'}=\left(2\pi\right)\left[-\left(2\pi\right)\cdot\alpha^{'}\cdot k^{'}+\beta^{'}\right]\textrm{ .}
\end{equation}
Comparing above two equations yields $N^{\left(k\right)}$,
\begin{equation}
N^{(k)}_{\textrm{IIR}}=-\left(2\cdot\pi\right)\cdot\alpha^{'}\cdot k^{'}+\beta^{'}\textrm{ .}
\end{equation}

\section{\label{sec:Design Criteria For the Sub-Band All-Pass Filters}Design Criteria For the Sub-Band All-Pass Filters}
We will use the same optimization framework as described in \cite{Munir2014} for the design of sub-band filter with the following modifications in the cost functions.

\subsection{Phase Transfer Cost Function, $\Psi_{\textrm{trans. phase}}^{\left(k\right)}$}
The objective is to design sub-band IIR all-pass equalizer whose phase response matches the desired phase response of (\ref{eq:Sub-Band Desired Phase Response}). Therefore, 
product of a sub-band all-pass equalizer and the sub-band CD channel has to be ideally 
\[
G_{\textrm{IIR}}^{\left(k\right)}\left(\omega^{'}\right)\cdot H_{\textrm{CD}}^{\left(k\right)}\left(\omega^{'}\right)=e^{-j\left(\phi_{0}^{\left(k\right)}+\beta^{'}\cdot\left(\omega^{'}-k^{'}\cdot\pi\right)\right)}\textrm{ .}\]
We will define error transfer function $\Upsilon^{\left(k\right)}\left(\omega\right)$ as
\[
\Upsilon^{\left(k\right)}\left(\omega^{'}\right)=G_{\textrm{IIR}}^{\left(k\right)}\left(\omega^{'}\right)\cdot H_{\textrm{CD}}^{\left(k\right)}\left(\omega^{'}\right)\cdot e^{j\left(\phi_{0}^{\left(k\right)}+\beta^{'}\cdot\left(\omega^{'}-k^{'}\cdot\pi\right)\right)}-1\textrm{ .}
\]
Weighted MSE in phase transfer can be written as
\[
\textrm{MSE}_{\textrm{trans. phase}}^{\left(k\right)}=\int_{-\pi}^{\pi}W\left(\omega^{'}\right)\cdot\left|\Upsilon^{\left(k\right)}\left(\omega^{'}\right)\right|^{2}\cdot d\omega^{'}\textrm{ .}\]
Therefore, optimization problem takes the form of minimizing the $\textrm{MSE}_{\textrm{trans. phase}}^{\left(k\right)}$ metric
\begin{equation}
\begin{aligned}
\Psi_{\textrm{trans. phase}}^{\left(k\right)}=&\min_{\rho_{i}^{\left(k\right)},\theta_{i}^{\left(k\right)},\phi_{0}^{\left(k\right)}}\textrm{MSE}_{\textrm{trans. phase}}^{\left(k\right)}  \\
&\textrm{ s.t. }\mid\rho_{i}^{\left(k\right)}\mid<1,\; i=1,2,\ldots,N^{(k)}_{\textrm{IIR}}\textrm{ .}\label{eq: Sub-Band 2nd Cost Function}
\end{aligned}
\end{equation}

\subsection{Group Delay Cost Function, $\Psi_{\textrm{GD}}^{\left(k\right)}$}
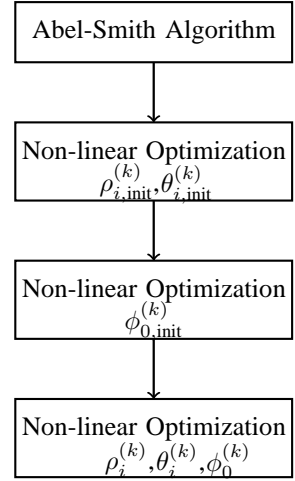
\begin{figure}

\centering{
\hspace{-0.5cm}
\begin{tikzpicture}[scale=0.80]
    
		
		\draw [style=thick] (-0.3,0-2) rectangle(4.3,-1);	 \node at (2,0.5-2) {Abel-Smith Algorithm};	
		
		\draw [->,line width=0.3mm] (2,0-2) -- (2,-1-2);
		\draw [style=thick] (-0.3,0-2-2.3) rectangle(4.3,-1-2);	 \node at (2,0.5-2-2) {Non-linear Optimization};	\node at (1.6,0.5-2-2-0.5) {$\rho^{(k)}_{i,{\textrm{init}}}\textrm{,}$};	\node at (2.55,0.5-2-2-0.53) {$\theta^{(k)}_{i,{ \textrm{init}}}$};	
		
		\draw [->,line width=0.3mm] (2,0-2-2.3) -- (2,-2-2.3-1);
		\draw [style=thick] (-0.3,-2-2.3-2.3) rectangle(4.3,-2-2.3-1);	 \node at (2,0.5-2-2-2.3) {Non-linear Optimization};	\node at (2,0.5-2-2-0.4-2.4) {$\phi^{(k)}_{0,{ \textrm{init}}}$};	
		
		\draw [->,line width=0.3mm] (2,0-2-2.3-2.3) -- (2,-2-2.3-1-2.3);
		\draw [style=thick] (-0.3,-2-2.3-2.3-2.3) rectangle(4.3,-2-2.3-1-2.3);	\node at (2,0.5-2-2-2.3-2.3) {Non-linear Optimization};	
		\node at (1.6,0.5-2-2-0.4-2.3-2.4) {$\rho^{(k)}_{i}\textrm{,}$};	\node at (2.45,0.5-2-2-0.4-2.3-2.42) {$\theta^{(k)}_{i}\textrm{,}$};	\node at (3.25,0.5-2-2-0.4-2.3-2.42) {$\phi^{(k)}_{0}$};
		
		
						
\end{tikzpicture}\medskip{}
\caption{\label{fig:block diagram of optimization framework} Block diagram of optimization framework.}
}
\end{figure}

Weighted MSE for the $k^{th}$ sub-band in terms of group delay can be written as
\begin{equation}
\textrm{MSE}_{\textrm{GD}}^{\left(k\right)}=\int_{-\pi}^{\pi}W\left(\omega^{'}\right)\cdot\mid\tau_{\textrm{Desired}}^{\left(k\right)}\left(\omega^{'}\right)-\sum_{i=1}^{N_{\textrm{IIR}}^{\left(k\right)}}\tau_{\textrm{IIR}_{i}}^{\left(k\right)}\left(\omega^{'}\right)\mid^{2}d\omega^{'}\textrm{ .}\label{eq:Sub-Band MSE in Group Delay}
\end{equation}
Therefore, optimization problem takes the form of minimizing the $\textrm{MSE}_{\textrm{GD}}^{\left(k\right)}$ metric
\begin{equation}
\Psi_{\textrm{GD}}^{\left(k\right)}=\min_{\rho_{i}^{\left(k\right)},\theta_{i}^{\left(k\right)}}\, \textrm{MSE}_{\textrm{GD}}^{\left(k\right)}\textrm{ s.t. }\mid\rho_{i}^{\left(k\right)}\mid<1,\; i=1,2,\ldots,N^{(k)}_{\textrm{IIR}}\textrm{ .}\label{eq: Sub-Band 1st Cost Function}
\end{equation}
Fig. \ref{fig:block diagram of optimization framework} depicts the flow chart of FB based IIR all-pass sub-band optimization framework. 

\section{Complexity Analysis} 
\label{sec:Complexity Analysis}
Complexity of the IIR equalizer in terms of real multiplications is given by 
\[
C_{\textrm{IIR}}=4\cdot\left(N_{\textrm{IIR}}+1\right)\textrm{ .}\]
If we implement IIR in the sub-band domain then the complexity reduces by a factor $\frac{2}{M}$ since each sub-band is running at a lower rate. But
at the same time, the overall filter order $N_{\textrm{IIR}}$ increases by an overlapping factor $\kappa$ which is 2 in this case. Therefore, complexity of IIR with the FB is given
by
\begin{equation}
C_{\textrm{IIR}}^{'}=4\cdot\frac{2}{M}\cdot\left(\kappa\cdot N_{\textrm{IIR}}+M\right)=\frac{8\cdot\kappa\cdot N_{\textrm{IIR}}}{M}+8.\end{equation}
Total complexity of the NMDFT FB (AFB and SFB) \cite{Karp1999} and IIR is given as
\begin{equation}
C_{\textrm{FB}+\textrm{IIR}}=4\log_{2}\left(M\right)-6+8\cdot K+\frac{8\cdot\kappa\cdot N_{\textrm{IIR}}}{M}\textrm{ .}\label{eq:Complexity of IIR and Filter Bank}
\end{equation}
In (\ref{eq:Complexity of IIR and Filter Bank}), the only variable is the number of sub-bands and minimum value of $C_{\textrm{FB}+\textrm{IIR}}$ is obtained
by setting the derivative of (\ref{eq:Complexity of IIR and Filter Bank}) with respect to $M$ 
to zero and solving it to find the optimal value of $M$, i.e.,
\begin{equation}
\therefore M_{\textrm{opt}} = 2\cdot\kappa\cdot N_{\textrm{IIR}}\cdot\ln\left(2\right)\textrm{ .}\label{eq:Optimum No of Subbands}
\end{equation}
Substituting $M_{\textrm{opt}}$ in (\ref{eq:Complexity of IIR and Filter Bank}), we will get
\begin{equation}
\left(C_{\textrm{FB}+\textrm{IIR}}\right)_{\textrm{opt}}=4\log_{2}\left(2\cdot\kappa\cdot N_{\textrm{IIR}}\cdot\ln\left(2\right)\right)-6+8\cdot K+\frac{4}{\ln\left(2\right)}\textrm{ ,}\label{eq:Optimium Complexity FB+IIR}\end{equation}
which increases logarithmically with $N_{\textrm{IIR}}$. 
\section{Simulations}
\label{sec:Simulations}
\begin{figure}
\centering\includegraphics[scale=0.48]{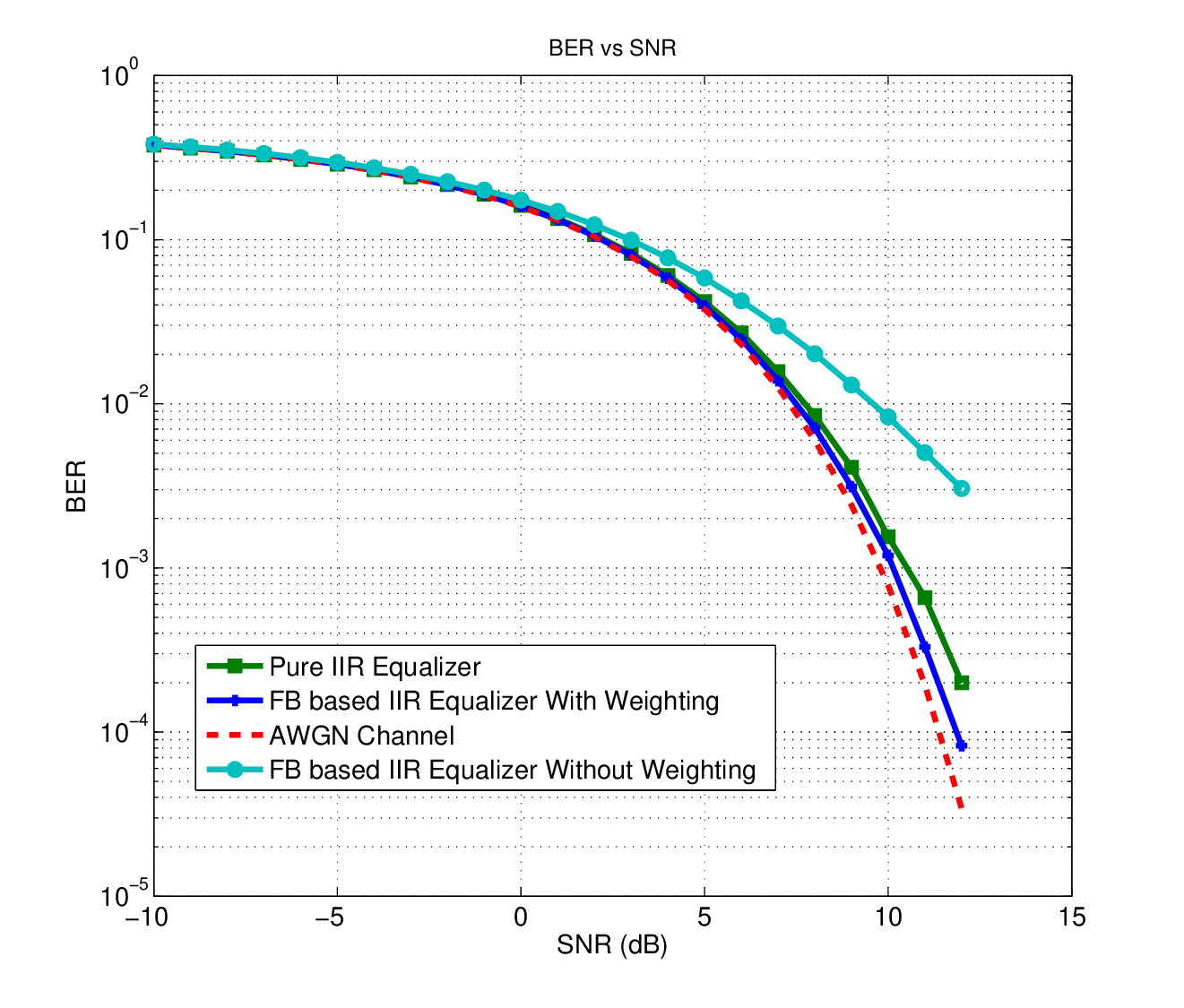}
\caption{\label{fig:BER Comp Between FBIIR With and Without Weighting}BER comparison between pure IIR equalization and 
filter bank based IIR equalization without and with weighting function. Simulations are performed with QPSK scheme for $M=32$, $K=8$, $L=2000\textrm{ km}$, squared magnitude RRC weighting function with $\omega_{c}^{'}=0.60\cdot\pi$ and roll-off is $0.1$.}
\end{figure}

A $14$ GBaud QPSK transmission with digital coherent receiver applying two-fold oversampling with $B=28$ GS/s is used to verify our technique for CD equalization. 
System parameters are $\lambda_0 =1550$ nm, $D=16$ ps/nm/km, $L=2000$ km, $M= 32$ and $K=8$. Same weighting function is used in the optimization framework since 
the prototype filters in all the sub-bands have same spectral shape. Fig. \ref{fig:BER Comp Between FBIIR With and Without Weighting} shows the BER comparison between FB 
based sub-band IIR equalization and pure IIR equalization. In the first set of simulation, FB based sub-band equalizer with uniform weighting in each sub-band is used and BER 
is calculated. It can seen from the figure that it has worse performance compared to pure IIR equalization. In the second set of simulations, FB based sub-band equalizer with 
squared magnitude RRC weighting function is used. BER simulations are carried out for different cut-off frequencies and roll-off factor. We found out the optimal parameters are $\omega_{c}^{'}=0.60\cdot\pi$ and 
roll-off is $0.1$ for squared magnitude RRC weighting function. Fig. \ref{fig:BER Comp Between FBIIR With and Without Weighting} shows the improvement with the weighting function and it slightly outperform the 
pure IIR equalization.

Complexity of a FB based processing depends on the length of non-trivial prototype filter and it is reflected in the parameter $K$ in (\ref{eq:Complexity of IIR and Filter Bank}). 
In our proposed FB structure, the prototype is the RRC filter with fixed cut-off frequency $\frac{\pi}{M}$ but roll-off is a design parameter. Length 
of prototype filter should be very large for small roll-off factor and vise versa. All the simulation till this point are reported for roll-off factor of $0.2$ and 
filter length of $256$. We are interested in reducing the length of prototype filter with very small loss in 
BER performance. Table \ref{tab:BER Performance for Diff Roll-Off and Filter Length} shows the effect of reducing the length for two cases. 
From these observations, we conclude that the roll-off factor of RRC filter for other values $M$ should also be takes as a design parameter in the optimization framework. 
\begin{table}[h]
\centering{}\begin{tabular}{|c|c|c|}
\hline 
\multicolumn{1}{|c|}{Roll-Off} & $L_{\textrm{RRC}}=K\cdot M$ & SNR at $10^{-3}$ BER\tabularnewline
\hline
\hline 
0.2 & 256 & 10.1\tabularnewline
\hline 
 & 128 & 11.2\tabularnewline
\hline 
0.9 & 256 & 10.1\tabularnewline
\hline 
 & 128 & 10.1\tabularnewline
\hline 
 & 64 & 10.2\tabularnewline
\hline
\end{tabular}
\medskip{}
\caption{\label{tab:BER Performance for Diff Roll-Off and Filter Length}\normalfont BER
performance for different roll-off and filter length of RRC protoype
filter. Number of sub-bands are chosen as $M=256$ for the simulations.}
\end{table}

\begin{figure}
\centering\includegraphics[scale=0.46]{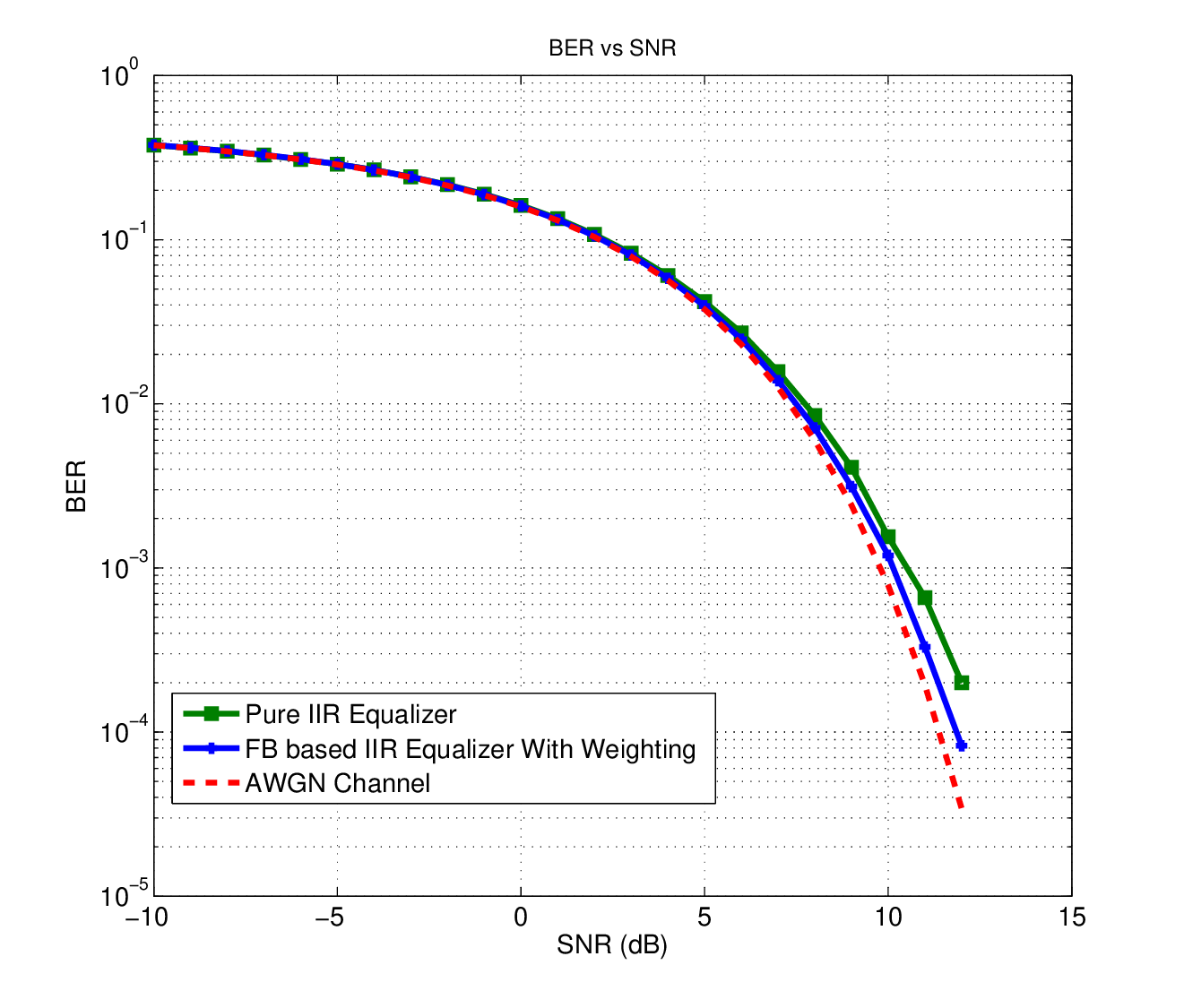}
\caption{\label{fig:BER Comp Between FBIIR With and Without Weighting256}BER comparison between pure IIR equalization and 
filter bank based IIR equalization without and with weighting function. Simulations are performed with QPSK scheme for $M=256$, $K=2$, $L=2000\textrm{ km}$, squared magnitude RRC weighting function with $\omega_{c}^{'}=0.60\cdot\pi$ and roll-off is $0.1$.}
\end{figure}

One of the major advantage of our methodology compared to the existing ones is the flexibilty in selecting the number of sub-bands $M$. Based on the simulation setting mentioned earlier, the optimum number of sub-band is $M_{\textrm{opt}} \approx 256$ according to (\ref{eq:Optimum No of Subbands}). Fig. \ref{fig:BER Comp Between FBIIR With and Without Weighting256} shows BER results for $M=256$ and it is almost the same as that of $M=32$. 

\section{Conclusions}
\label{sec:Conclusions}
We presented NMDFT FB based framework with RRC prototype filters in the AFB and SFB to design sub-band IIR all-pass filters for CD compensation. 
Necessary conditions based on group delay characteristic are derived to select minimum filter order and stability of the resulting sub-band IIR all-pass
filters. An optimization scheme presented in \cite{Munir2014} is used to find the coefficients of the sub-band IIR all-pass filters. Very long optical links 
can be compensated by this method as the complexity increases logarithmically with the fiber length. Simulation result show 
that BER performance of the proposed method with appropriate selection of weighting function in the optimization framework is better compared to pure IIR 
equalization. Additionally we show that a compromise between complexity and performance can be reached by selecting roll-off factors and filter lengths of 
the RRC prototype filter. One of the advantage offered by our scheme over FDE is the flexibility in selecting $M$ which might be attractive from 
hardware implementation point of view. It is left as a future work to compare the complexity and performance comparison between FDE equalization and our proposed scheme. 




%

\bibliographystyle{IEEEtran}

\end{document}